\documentclass[british,english,aps,manuscript,superscriptaddress,manuscript]{revtex4}
\usepackage[T1]{fontenc}
\usepackage[latin9]{inputenc}
\usepackage{babel}
\usepackage{array}
\usepackage{verbatim}
\usepackage{float}
\usepackage{multirow}
\usepackage{amsmath}
\usepackage{graphicx}
\usepackage[unicode=true,
 bookmarks=true,bookmarksnumbered=false,bookmarksopen=false,
 breaklinks=false,pdfborder={0 0 1},backref=false,colorlinks=false]
 {hyperref}
\hypersetup{pdftitle={Kinetic Theory of Positron-Impact Ionization in Gases},
 pdfauthor={G. J. Boyle, W. J. Tattersall, D. G. Cocks, S. Dujko, R. D. White}}

\makeatletter

\newcommand{\lyxmathsym}[1]{\ifmmode\begingroup\def\b@ld{bold}
  \text{\ifx\math@version\b@ld\bfseries\fi#1}\endgroup\else#1\fi}

\providecommand{\tabularnewline}{\\}

\@ifundefined{textcolor}{}
{%
 \definecolor{BLACK}{gray}{0}
 \definecolor{WHITE}{gray}{1}
 \definecolor{RED}{rgb}{1,0,0}
 \definecolor{GREEN}{rgb}{0,1,0}
 \definecolor{BLUE}{rgb}{0,0,1}
 \definecolor{CYAN}{cmyk}{1,0,0,0}
 \definecolor{MAGENTA}{cmyk}{0,1,0,0}
 \definecolor{YELLOW}{cmyk}{0,0,1,0}
}

\makeatother

\begin{document}

\title{Kinetic Theory of Positron-Impact Ionization in Gases}

\author{G. J. Boyle}

\address{College of Science, Technology \& Engineering, James Cook University,
Townsville 4810, Australia}

\author{W. J. Tattersall}

\affiliation{Research School of Physics and Engineering, The Australian National
University, Canberra, ACT 0200, Australia}

\address{College of Science, Technology \& Engineering, James Cook University,
Townsville 4810, Australia}

\author{D. G. Cocks}

\address{College of Science, Technology \& Engineering, James Cook University,
Townsville 4810, Australia}

\author{S. Dujko}

\address{Institute of Physics, University of Belgrade, Pregrevica 118, 11080
Belgrade, Serbia}

\author{R. D. White}

\address{College of Science, Technology \& Engineering, James Cook University,
Townsville 4810, Australia}
\begin{abstract}
A kinetic theory model is developed for positron-impact ionization
(PII) with neutral, rarefied gases. Particular attention is given
to the sharing of available energy between the post-ionization constituents.
A simple model for the energy-partition function that qualitatively
captures the physics of high-energy and near-threshold ionization
is developed for PII, with free parameters that can be used to fit
the model to experimental data. By applying the model to the measurements
of Kover and Laricchia \cite{KoveLari98} for positrons in H$_{2}$,
the role of energy-partitioning in PII for positron thermalisation
is studied. Although the overall thermalisation time is found to be
relatively insensitive to the energy-partitioning, the mean energy
profiles at certain times can differ by more than an order of magnitude
for the various treatments of energy-parititioning. This can significantly
impact the number and energy distribution of secondary electrons.
\end{abstract}

\keywords{Positrons, Positron impact ionization, Kinetic theory, Monte Carlo,
Molecular Hydrogen}

\maketitle

\section{Introduction\label{sec:Introduction}}

An understanding of the behavior of positrons in gases underpins many
areas of technology and scientific research \cite{MurpSurk92,Greaetal94,Cole02,Petretal13,Pano13}.
Of particular interest are applications to the medical imaging technique
of Positron Emission Tomography (PET) \cite{CzerPhel02}. To optimize
PET technologies and quantify the associated radiation damage requires
a thorough understanding of the processes by which an energetic positron
(and the secondary species) thermalise. It has been shown recently
by Sanche and co-workers \cite{Sanc03,Sanc05,Sanc08,Sanc09} that
the secondary electrons created via ionization can cause significant
DNA damage. The number of secondary electrons ejected along the positron
track is on the order of $10^{4}$ per MeV of primary radiation produced
in water \cite{Abrietal11,Cobuetal97}, so it is clear that particular
attention needs to be paid to the ionization process. 

Although there has been extensive research on electrons in gases,
positrons remain significantly less well understood. Specific collisional
processes are avilable to the positron which do not exist for electrons,
e.g. annihilation with an electron and positronium formation \cite{CharHumb01,Kaup86}.
Although the impact from either a sufficiently energetic positron
or electron can ionize a gas molecule, the ionization process differs
in a crucial way; ionization by positron impact is a particle-conserving
process with respect to positrons, while ionization by electron impact
is non-particle-conserving with respect to electrons \cite{LoffWink96,NessNolan00,Winketal02,GrubLoff14,Dujketal14}.
The two types of ionization will be referred to as `positron-impact
ionization' (PII) and `electron-impact ionization' (EII) respectively.
In the framework of kinetic theory, Ness \cite{Ness85} developed
a collision operator for EII but no positron equivalent has yet been
developed. Instead, previous investigations \cite{Suvaetal08,Marletal09,WhitRobs09,WhitRobs11,Banketal12,Banketal12b}
have generally treated positron ionization as a simple excitation
process which effectively assumes that the scattered positron receives
all of the available post-ionization energy, although \cite{Petretal14}
has highlighted the effects of the secondary electron energy distribution.

In this paper, a PII equivalent of the EII collision operator of Ness
is derived for the first time. Macroscopic transport coefficients,
such as mean energy and flux drift velocity, are compared for a simple
benchmark model using both a kinetic theory approach based on the
Boltzmann equation, and Monte Carlo simulation. Particular attention
is paid to the effect of  energy-sharing between post-ionization constituents,
and the influence that different energy-partitioning models have on
transport. A basic energy-partitioning model that captures, at least
qualitatively, the physics of high energy and near-threshold positron
ionization is proposed, which can then be fitted to the rather limited
experimental data that is available. The new kinetic theory model
is used to investigate the transport of positrons in dilute H$_{2}$
gas using a recently-compiled complete set of cross sections \cite{Petretal13},
and the proposed energy-partitioning model fitted to the experimental
data of Kover and Laricchia \cite{KoveLari98}.

\section{Theory\label{sec:Theory}}

\subsection{The kinetic equation and its multi-term solution\label{sub:The-kinetic-equation}}

The fundamental equation describing a swarm of positrons moving through
a dilute gaseous medium subject to an electric field, $\mathbf{E}$,
is the Boltzmann kinetic equation for the phase-space distribution
function $f\equiv f\left(\mathbf{r},\mathbf{v},t\right)$ \cite{WhitRobs11}:

\begin{align}
\left(\frac{\partial}{\partial t}+\mathbf{v}\cdot\nabla+\frac{q\mathbf{E}}{m}\cdot\frac{\partial}{\partial\mathbf{v}}\right)f & =-J\left(f\right),\label{eq:Boltz}
\end{align}
where $t$ is the time, and $\mathbf{r}$, $\mathbf{v}$, $q$ and
$m$ are the position, velocity, charge and mass of the positron respectively.
The right hand side describes the effect of collisions on the distribution
function at a fixed position and velocity. Essentially, the Boltzmann
equation is an equation of continuity in phase-space \cite{Cerc98}.
Solving equation (\ref{eq:Boltz}) for the distribution function yields
all relevant information about the system. Macroscopic transport properties
including mean energy and drift velocity can then be found via averages
over the ensemble as detailed in Section \ref{sub:Transport-properties}.
The purpose of this paper is to investigate the effect of ionization,
so for simplicity we will consider only spatially-homogeneous situations.

If there is a single preferred direction in the system, e.g. due to
an electric field in plane parallel geometry, then the angular dependence
of the velocity component can be adequately described by an expansion
in terms of Legendre polynomials \cite{RobsWinkSige02}, i.e. if $f\left(\mathbf{v},t\right)\rightarrow f\left(v,\mu,t\right)$,
where $\mu=\hat{\mathbf{v}}\cdot\mathbf{\hat{E}}$, then 
\begin{align}
f\left(\mathbf{v},t\right) & =\sum_{l=0}^{\infty}f_{l}\left(v,t\right)P_{l}(\mu),\label{eq:Legendre}
\end{align}
where $P_{l}$ is the $l$-th Legendre polynomial \cite{AbraSteg72}.
Substituting the expansion~(\ref{eq:Legendre}) into equation~(\ref{eq:Boltz})
and equating the coefficients of Legendre polynomials results in the
following coupled partial differential equations for the $f_{l}$
 in energy-space, 

\begin{align}
\frac{\partial f_{l}}{\partial t} & +\sum_{p=\pm1}\Delta_{l}^{(p)}\frac{qE}{m}\left(U^{^{\frac{1}{2}}}\frac{\partial}{\partial U}+p\frac{\left(l+\frac{3p+1}{2}\right)}{2}U^{^{-\frac{1}{2}}}\right)f_{l+p}\begin{array}[t]{cc}
=-J_{l}\left(f_{l}\right) & \left(l=0,1,2,\dots,\infty\right)\end{array},\label{eq:BoltzLeg}
\end{align}
where $U=\frac{1}{2}mv^{2}$, $J_{l}$ is the Legendre decomposition
of the collision operator, and
\begin{align*}
\Delta_{l}^{(+1)} & =\frac{(l+1)}{(2l+3)},\\
\Delta_{l}^{(-1)} & =\frac{l}{(2l-1)}.
\end{align*}

Equation (\ref{eq:BoltzLeg}) represents an infinite set of coupled
partial differential equations for the expansion coefficients, $f_{l}$.
In practice, one must truncate the series (\ref{eq:Legendre}) at
a sufficiently high index, $l=l_{\mathrm{max}}$. The history of charged
particle transport in gases has been dominated by the `two-term approximation'
\cite{HuxlCrom74}, i.e., where only the first two terms have been
included. The assumption of quasi-isotropy necessary for the two-term
approximation is violated in many situations, particularly when inelastic
collisions are included \cite{Whitetal02} or when higher order moments
are probed \cite{Boyle15}. Such an assumption is not necessary in
our formalism. Rather, $l_{\mathrm{max}}$ is treated as a free parameter
to be increased until some convergence or accuracy criterion is met.

\subsection{Collision operators in the multi-term representation\label{sub:Collision-operators}}

To solve equation (\ref{eq:BoltzLeg}) we require the collision operators
for all of the relevant collisional processes, and their representations
in terms of Legendre polynomials, $J_{l}$. If we assume that the
neutral background gas is at rest and in thermal equilibrium at a
temperature $T_{0}$, then the background medium has a Maxwellian
distribution in velocity space and the collision operator is linear
in the swarm approximation \cite{Robs06}. Below we detail the specific
kinetic theory forms of the collision operator for conservative elastic
and inelastic collisions, particle-loss collisions such as annihilation
and positronium formation, and ionization, which is the focus of this
work. A further expansion of each collision integral with respect
to the ratio of swarm particle mass to neutral particle mass, $m/m_{0}$,
has been performed. Because this ratio is small for positrons (and
electrons), only the leading term of this expansion for each collision
process and in each equation of the system (\ref{eq:BoltzLeg}) was
taken into account.

The total collision operator can then be separated for each of the
different types of processes, eg.
\begin{align*}
J & =J^{\mathrm{el}}+J^{\mathrm{in}}+J^{\mathrm{ann}}+J^{\mathrm{Ps}}+J^{\mathrm{ion}},
\end{align*}
where the right hand side terms represent the elastic, inelastic,
annihilation, positronium formation and ionization collision operators
respectively. Microscopic scattering information is included via the
appropriate scattering cross sections \cite{GoldPoolSafk01,KaupSteiWade85}.
It is more natural to work with the collision frequency rather than
the scattering cross sections directly. A collision frequency, $\nu$,
is defined for a particular process by
\begin{align}
\nu(U) & \equiv n_{0}\sqrt{\frac{2}{m}}U^{\frac{1}{2}}\sigma(U),\label{eq:colfreq}
\end{align}
where $\sigma$ is the corresponding cross section of the process.

\subsubsection{Conservative elastic and inelastic collisions\label{sub:Conservative}}

For particle-conserving elastic and inelastic collisions we assume
the Wang-Chang et al.~\cite{WangChanetal64} semi-classical collision
operator and its limiting cases. For an elastic collision, if all
terms proportional to the mass ratio are neglected there is no energy
transfer during a collision. To obtain a non-zero expression, a first-order
mass ratio approximation is required \cite{Davy35}, i.e. 
\[
J_{l}^{\mathrm{el}}\left(f_{l}\right)=\begin{cases}
-\frac{2m}{m_{0}}U^{-\frac{1}{2}}\frac{\partial}{\partial U}\left[U^{\frac{3}{2}}\nu_{1}^{\mathrm{el}}(U)\left(f_{0}+kT_{0}\frac{\partial f_{0}}{\partial U}\right)\right] & l=0,\\
\nu_{l}^{\mathrm{el}}(U)f_{l} & l\geq1,
\end{cases}
\]
where $\nu_{l}^{\mathrm{el}}=n_{0}\sqrt{\frac{2}{m}}U^{\frac{1}{2}}\left(\sigma_{0}^{\mathrm{el}}-\sigma_{l}^{\mathrm{el}}\right)$,
and $\sigma_{l}$ is defined from the differential scattering cross
section \cite{GoldPoolSafk01}, $\sigma(U,\mu)$, via,
\[
\sigma_{l}\left(U\right)=2\pi\int_{0}^{\pi}d\mu P_{l}\left(\mu\right)\sigma(U,\mu).
\]

If the background gas has internal degrees of freedom then, to zeroth
order in the mass ratio, energy exchange can still occur through excitation
and de-excitation of those internal states. Hence, unlike the isotropic
part of the elastic collision integral, the scalar part of the inelastic
collision integral does not vanish under a zeroth order mass assumption.
The Legendre decomposed form of the inelastic collision operator in
the cold gas limit is given by \cite{FrostPhelp62,Hols46}
\begin{align}
J_{l}^{\mathrm{in}}\left(f_{l}\right) & =\sum_{j}\nu_{j}^{\mathrm{in}}\left(U\right)f_{l}-\begin{cases}
\begin{array}{c}
\left(\frac{U+U_{j}}{U}\right)^{\frac{1}{2}}\nu_{j}^{\mathrm{in}}\left(U+U_{j}\right)f_{l}\left(U+U_{j}\right)\\
0
\end{array} & \begin{array}{c}
l=0,\\
l\geq1,
\end{array}\end{cases}\label{eq:inelastic}
\end{align}
where the subscript $j$ denotes the available inelastic channels,
such as excitations and rotations, with an associated inelastic scattering
cross section $\sigma_{j}^{\mathrm{in}}(U)$, and a threshold energy
$U_{j}$. It is implicit in the above equation that there is no thermal
excitation of internal states.

\subsubsection{Annihilation and positronium formation\label{sub:Loss}}

Positron annihilation and positronium formation occur through distinctly
different physical mechanisms. However, from a transport theory perspective
they each represent a unidirectional particle loss process, and hence
the form of their collision operators are identical. Since there is
no post-collision scattering the collision operator is simply \cite{NessRobs86},
\begin{align*}
J_{l}^{\mathrm{loss}}\left(f_{l}\right) & =\sum_{k}\nu_{k}^{\mathrm{loss}}(U)f_{l},
\end{align*}
where $k$ are the available loss process channels, and $\nu_{k}^{\mathrm{loss}}$
is the collision frequency for the $k$th loss process corresponding
to the cross section $\sigma_{k}^{\mathrm{loss}}(U)$.

\subsubsection{Ionization\label{sub:Ionization}}

Ionization by electron impact is fundamentally different from ionization
by positron impact. Since the ejected electron is of the same species
as the impacting particle, EII is a non-particle-conserving process,
i.e., the indistinguishability of electrons leads to a gain in the
number of electrons in the swarm. Since the scattered positron can
be distinguished from the ejected electron, PII is a particle-conserving
process. A different collision operator needs to be used for each
case.  In previous studies, PII was treated as a simple excitation
process, which ignores the possible partitioning of energy between
the scattered positron and ejected electron. In what follows, we develop
an explicit expression for the PII operator.

Following the approach of \cite{Ness85}, the details of which are
given in Appendix~\ref{sub:DerivationIonisation}, the PII collision
operator takes the form 
\begin{align}
J_{l}^{\mathrm{ion}}\left(f_{l}\right) & =\nu^{\mathrm{ion}}\left(U\right)f_{l}\left(U\right)-\begin{cases}
\begin{array}{c}
{\displaystyle \int}\left(\frac{U'}{U}\right)^{^{\frac{1}{2}}}P(U,U')\nu^{\mathrm{ion}}\left(U'\right)f_{0}(U')dU'\\
0
\end{array} & \begin{array}{c}
l=0,\\
l\geq1,
\end{array}\end{cases}\label{eq:IonC}
\end{align}
where $U'$ is the impact particle energy, and $\nu^{\mathrm{ion}}$
is the collision frequency for ionization, corresponding to an ionization
cross section. The $P(U,U')$ term is the energy-partitioning function,
defined such that $P(U,U')dU$ represents the probability of the positron
having an energy in the range $U+dU$ for an incident positron of
energy $U'$. The energy-partitioning function has the following properties:
\begin{align*}
P(U,U') & =\begin{array}{cc}
0\ \mbox{for}\  & U'<U+U_{I},\end{array}\\
\int_{0}^{U'-U_{I}}\negthickspace P(U,U')dU & =\begin{array}{cc}
1\ \mbox{for}\  & U'\geq U+U_{I},\end{array}
\end{align*}
where $U_{I}$ is the ionization threshold energy, i.e., the energy
needed to overcome the electron binding. The energy-sharing, which
is determined by the energy-partitioning function $P$, is a major
theme in the present work. It will be shown in Section \ref{sec:Results-and-discussion}
that different energy-partition models significantly affect positron
transport.

\subsection{Transport properties\label{sub:Transport-properties}}

The cross sections and collision operator terms represent the microscopic
picture of positron interactions with the medium. The macroscopic
picture, e.g. transport properties that represent experimental measurables,
are obtained as averages of certain quantities with respect to the
distribution function, $f$. Among the transport properties of interest
in the current manuscript are the number density, $n$, flux drift
velocity, $W$, and mean energy, $\epsilon$, of the positron swarm,
which can be calculated via \cite{Robs06}
\begin{align*}
n & =2\pi\left(\frac{2}{m}\right)^{\frac{3}{2}}\int dU\ U^{\frac{1}{2}}f_{0}(U),\\
W & =\frac{1}{n}\frac{2\pi}{3}\left(\frac{2}{m}\right)^{2}\int dU\ Uf_{1}(U),\\
\epsilon & =\frac{1}{n}2\pi\left(\frac{2}{m}\right)^{\frac{3}{2}}\int dU\ U^{\frac{3}{2}}f_{0}(U).
\end{align*}
The focus of this paper is the ionization process, so it is also useful
to calculate the average ionization collision rate defined by
\begin{align*}
\alpha^{\mathrm{ion}} & =\frac{1}{n}2\pi\left(\frac{2}{m}\right)^{2}\int dU\ U^{\frac{1}{2}}\nu^{\mathrm{ion}}(U)f_{0}(U).
\end{align*}

\section{The numerical approach for a multi-term solution \label{sec:Numerical-approach}}

In this section we detail a numerical solution of the system of coupled
ordinary differential equations, (\ref{eq:BoltzLeg}), once an $l$-index
truncation has been applied.

\subsection{Method of lines}

The Method of Lines (MOL) \cite{Sham05,SadiObio00} is a technique
for solving PDEs in which all but one dimension is discretized. In
developing a numerical solution to the Boltzmann equation, we choose
to first discretize the energy- (or equivalently, speed-) space. In
general, applying the MOL to linear PDEs results in a system of equations
of the form
\begin{align}
\mathbf{M}\frac{d}{dt}\mathbf{u} & =\mathbf{L}\mathbf{u},\label{eq:PDEmassmatrix}
\end{align}
where $\left[u\right]_{i}=u_{i}(t)\equiv u(x_{i},t),$ and $\mathbf{L}$
and $\mathbf{M}$ are matrices resulting from the discretization process,
commonly known as the ``Stiffness Matrix'' and ``Mass Matrix''
respectively \cite{CarvHind78}. The formerly continuous variable
$x$ has been discretized into a set of $x_{i}$ for $i=0,1,...,n.$
The MOL formalism allows easy implementation of linear boundary conditions
or constraints via the mass matrix. Let the discretized boundary conditions
and constraints of (\ref{eq:PDEmassmatrix}) be represented by $\mathbf{G}\mathbf{u}=\mathbf{0}$,
where $\mathbf{G}$ is a matrix and $\mathbf{0}$ represents a vector
of zeros. Then clearly $\frac{d}{dt}\mathbf{G}\mathbf{u}=\mathbf{G}\frac{d}{dt}\mathbf{u}=\mathbf{0}$
and, provided the initial solution satisfies the constraints,
\begin{align}
\mathbf{\overline{M}}\frac{d}{dt}\mathbf{u} & =\overline{\mathbf{L}}\mathbf{u},\label{eq:PDEmassmatrix2}
\end{align}
where $\mathbf{\overline{M}}$ and $\overline{\mathbf{L}}$ are the
modified mass and stiffness matrices, 
\begin{align*}
\mathbf{\overline{M}}=\left[\begin{array}{c}
\mathbf{G}\\
\mathbf{M}
\end{array}\right] & ,\;\overline{\mathbf{L}}=\left[\begin{array}{c}
\mathbf{0}\\
\mathbf{L}
\end{array}\right].
\end{align*}

In a pure MOL approach, the system of ODEs, (\ref{eq:PDEmassmatrix2}),
are solved analytically. However, one is eventually forced to discretize
the time variable as well for complicated systems of equations, such
as those arising from the discretization of the Boltzmann equation.
In this work we choose to discretize the time dimension with a first-order
implicit Euler method~\foreignlanguage{british}{\cite{Butc03}},
for its good stability properties. Applying the implicit Euler method
to equation (\ref{eq:PDEmassmatrix}) or (\ref{eq:PDEmassmatrix2})
gives
\begin{align}
\left(\mathbf{M}-h\mathbf{L}\right)\mathbf{u}^{n+1} & =\mathbf{M}\mathbf{u}^{n}\label{eq:Euler}
\end{align}
where $\mathbf{u}^{n}$ and $\mathbf{u}^{n+1}$ are the solution vector,
$\mathbf{u}$, at times $t_{n}$ and $t_{n+1}$, and $h=t_{n+1}-t_{n}$
is the time step. For linear systems, equation (\ref{eq:Euler}) can
be solved directly with linear algebra techniques.

\subsection{Finite difference representation in energy-space\label{sub:Finite-Difference-representation}}

The finite difference method \cite{BurdFair93} is a local approximation
method which seeks to replace the continuous derivatives by a weighted
difference-quotient of neighboring points. It is widely used, simple
to program, and leads to sparse matrices with band structures approximating
derivatives \cite{Leve07}. Similar to the work of Winkler and collaborators
\cite{Winketal84,LoffWink96,Leyhetal98}, the system of ODEs is discretized
at centered points using a centered difference scheme, i.e.,
\begin{align*}
\left.\frac{df(U,t)}{dx}\right|_{U_{i+1/2}} & =\frac{f(U_{i+1},t)-f(U_{i},t)}{U_{i+1}-U_{i}},\\
f(U_{i+1/2},t) & =\frac{f(U_{i+1})+f(U_{i})}{2}.
\end{align*}
Although a general form can be constructed for an arbitrary grid,
the simplest case is for evenly spaced points, i.e. 
\[
U_{i}=i\Delta U\ \ 0\leq i\leq n,
\]
where $\Delta U$ is a constant. Discretizing at the center between
two solution nodes results in a system of linear equations that is
underdetermined. The extra information is naturally provided by boundary
conditions which are appended to the system.

\subsection{Initial and boundary conditions}

In positron experiments \cite{CharHumb01}, unmoderated positrons
have a peak in their emission energy spectrum of around half an MeV,
which then lose energy rapidly via collisions. There is little information
about the initial source distribution in thermalisation experiments
\cite{CampHumb77}. For our purposes, we wish to probe the influence
of PII collisions, and accordingly choose an initial distribution
with a mean energy far above the ionization threshold so that a large
range of the ionization cross section can be sampled during relaxation.
One of the source distributions used by Campeanu and Humberston \cite{CampHumb77}
in their investigations of helium is a distribution that is constant
in speed space up to some sufficiently high cut-off value, $v_{\mathrm{max}}=\sqrt{2U_{max}/m}$,
i.e. $f_{0}(v)=\Theta\left(v_{\mathrm{max}}-v\right)C$, where $\Theta(x)$
is the Heaviside step function. The mean energy of this distribution
function is given by $\epsilon=\frac{3}{5}U_{\mathrm{max}}$. We will
use this type of initial distribution for our investigations of thermalisation
and shall choose $U_{\mathrm{max}}$ to be sufficiently high to sample
the ionization cross sections accordingly. 

The system of coupled equations (\ref{eq:BoltzLeg}) requires boundary
conditions on the expansion coefficients $f_{l}$. Winkler and collaborators
\cite{Winketal84,LoffWink96,Leyhetal98} have analyzed the multi-term,
even-order approximation, and discovered that the general solution
of the steady-state hierarchy contains $\tfrac{1}{2}\left(l_{\mathrm{max}}+1\right)$
non-singular and $\tfrac{1}{2}\left(l_{\mathrm{max}}+1\right)$ singular
fundamental solutions when $U$ approaches infinity, and the physically
relevant solution has to be sought within the non-singular part. They
give the boundary conditions necessary for the determination of the
non-singular physically relevant solution as

\begin{align*}
f_{l}(U=0) & =\begin{array}{cc}
0 & \mbox{for odd \ensuremath{l}},\end{array}\\
f_{l}(U=U_{\infty}) & =\begin{array}{cc}
0 & \mbox{for even \ensuremath{l}},\end{array}\\
f_{l}(U>U_{\infty}) & =\begin{array}{cc}
0 & \mbox{for all \ensuremath{l}},\end{array}
\end{align*}
where $U_{\infty}$ represents a sufficiently large energy. In practice,
$U_{\infty}$ has to be determined in a prior calculation, and is
chosen such that the value of $f_{0}(U_{\infty})$ is less than $10^{-10}$
of the maximum value of $f_{0}$.

\section{Results and discussion\label{sec:Results-and-discussion}}

In this section we will apply both the kinetic theory technique detailed
in the previous sections, and a Monte Carlo simulation \cite{Tattetal15},
to describe positron transport in a benchmark model, and positron
transport in real $H_{2}$ gas. Comparisons are made to EII where
possible. Particular attention is given to the role of energy-partitioning
between the scattered and ejected particles post-ionization, and a
simple energy-partitioning model is proposed to capture the underlying
physics.

\subsection{Positron ionization benchmarking}

We first discuss several benchmark models for EII which can act as
a test bed for our numerical techniques and solution model. The Lucas-Saelee
\cite{LucaSael75} model is a popular benchmark, but focuses on the
differences between excitation and ionization rather than energy-partitioning
specifically. Taniguchi et al. \cite{Tanietal77} modified the partition
function of the Lucas-Saelee model, which assumes a distribution with
all energy-sharing fractions equiprobable, to instead share energy
equally between the two electrons, but found that it did not alter
the transport coefficients significantly. Instead, Ness and Robson
\cite{NessRobs86} proposed a step model for testing energy-sharing
for EII, which was shown to have some variation for the partitionings
they investigated. The details of the model are:

\begin{align}
\sigma_{0}^{\mathrm{el}}-\sigma_{l}^{\mathrm{el}} & =10\ \mbox{\AA}^{2},\label{eq:testmodel}\\
\sigma^{\mathrm{in}} & =\begin{cases}
\begin{array}{c}
1\ \mbox{\ensuremath{\lyxmathsym{\AA}^{2}}}\\
0
\end{array} & \begin{array}{c}
U\geq10\ \mbox{eV},\\
U<10\ \mbox{eV},
\end{array}\end{cases}\nonumber \\
\sigma^{\mathrm{ion}} & =\begin{cases}
\begin{array}{c}
1\ \mbox{\AA}^{2}\\
0
\end{array} & \begin{array}{c}
U\geq15\ \mbox{eV},\\
U<15\ \mbox{eV},
\end{array}\end{cases}\nonumber \\
m_{0} & =25\ \mbox{amu},\nonumber \\
T_{0} & =0\ \mbox{K}.\nonumber 
\end{align}
Transport coefficients for EII calculated using kinetic theory are
compared against the results of Ness and Robson, and the Monte Carlo
simulations in Table \ref{tab:ElectronBenchmark} of Appendix \ref{sub:Electron-Impact-Ionization}.
The results support the integrity of our methods and solutions. Transport
coefficients for PII under this model are given in Table \ref{tab:Benchmark1}
for varying energy sharing fractions, $Q$, where $Q=\frac{U}{U'-U_{I}}$.
As described in Appendix \ref{sub:DerivationIonisation} the collision
operator~(\ref{eq:IonC}) breaks down when $Q=0$, hence there is
no value given in Table \ref{tab:Benchmark1} corresponding to the
kinetic model for positrons with $Q=0$. No previous positron impact
calculations exist for model (\ref{eq:testmodel}), so the transport
properties from our kinetic theory model are compared solely against
an independent Monte Carlo simulation in Table \ref{tab:Benchmark1}.
The uncertainty in the Monte Carlo simulations has been estimated
to be less than $1\%$ for the ionization collision rates, and less
than $0.5\%$ (generally less than $0.3\%$) for the drift velocity
and mean energy. The two approaches give $\alpha^{\mathrm{ion}}/n_{0}$,
$\epsilon$ and $W$ values which differ by less than $0.6\%$, $0.3\%$
and $0.3\%$ respectively, over the range of reduced electric fields
and available energy fractions, all of which are within the corresponding
Monte Carlo uncertainty.

As the reduced field, $E/n_{0}$, is increased, the velocity distribution
function samples more of the ionization process leading to a greater
ionization rate and a stronger dependence of the transport coefficients
on the post-collision energy partitioning as shown in Table \ref{tab:Benchmark1}.

\begin{table}[tph]
\protect\caption{\label{tab:Benchmark1}Comparison of average ionization rates, $\alpha^{\mathrm{ion}}/n_{0}$,
mean energies, $\epsilon$, and flux drift velocities, $W$, for PII
for model (\ref{eq:testmodel}) for different reduced fields $E/n_{0}$
and energy sharing fractions $Q$. Columns `Current' correspond to
the current kinetic theory calculations, and columns `MC' are the
results of the Monte Carlo simulation. Note, $Q=\mathrm{AFE}$ corresponds
to ``all fractions equiprobable''.}

\centering{}%
\begin{tabular}{|c|c|c|>{\centering}p{1.2cm}|c|>{\centering}p{1.2cm}|c|>{\centering}p{1.2cm}|}
\hline 
$E/n_{0}$ &  & \multicolumn{2}{c|}{$\alpha^{\mathrm{ion}}/n_{0}$} & \multicolumn{2}{c|}{$\epsilon$} & \multicolumn{2}{c|}{$W$}\tabularnewline
(Td) & $Q$ & \multicolumn{2}{c|}{($10^{-15}$m$^{3}$s$^{-1}$)} & \multicolumn{2}{c|}{(eV)} & \multicolumn{2}{c|}{($10^{5}$ms$^{-1}$)}\tabularnewline
\hline 
 &  & Current & MC & Current & MC & Current & MC\tabularnewline
\hline 
\hline 
$300$ & $0$ &  & 1.711 &  & 6.869 &  & 2.767\tabularnewline
\hline 
 & $1/4$ & 1.720 & 1.718 & 6.919 & 6.931 & 2.722 & 2.730\tabularnewline
\hline 
 & $1/3$ & 1.725 & 1.719 & 6.940 & 6.942 & 2.711 & 2.706\tabularnewline
\hline 
 & $1/2$ & 1.740 & 1.739 & 6.983 & 6.979 & 2.693 & 2.689\tabularnewline
\hline 
 & $2/3$ & 1.757 & 1.761 & 7.021 & 7.023 & 2.677 & 2.676\tabularnewline
\hline 
 & $3/4$ & 1.767 & 1.774 & 7.041 & 7.040 & 2.671 & 2.664\tabularnewline
\hline 
 & $1$ & 1.807 & 1.804 & 7.098 & 7.087 & 2.654 & 2.648\tabularnewline
\hline 
 & AFE & 1.745 & 1.739 & 6.979 & 6.981 & 2.699 & 2.701\tabularnewline
\hline 
$500$ & $0$ &  & 4.856 &  & 9.210 &  & 3.951\tabularnewline
\hline 
 & $1/4$ & 4.915 & 4.917 & 9.379 & 9.375 & 3.819 & 3.822\tabularnewline
\hline 
 & $1/3$ & 4.955 & 4.949 & 9.446 & 9.450 & 3.789 & 3.780\tabularnewline
\hline 
 & $1/2$ & 5.060 & 5.055 & 9.579 & 9.588 & 3.738 & 3.739\tabularnewline
\hline 
 & $2/3$ & 5.211 & 5.208 & 9.716 & 9.714 & 3.697 & 3.697\tabularnewline
\hline 
 & $3/4$ & 5.288 & 5.293 & 9.788 & 9.789 & 3.678 & 3.678\tabularnewline
\hline 
 & $1$ & 5.565 & 5.599 & 10.03 & 10.05 & 3.627 & 3.628\tabularnewline
\hline 
 & AFE & 5.119 & 5.107 & 9.589 & 9.577 & 3.754 & 3.755\tabularnewline
\hline 
$800$ & $0$ &  & 9.903 &  & 13.30 &  & 5.260\tabularnewline
\hline 
 & $1/4$ & 10.21 & 10.23 & 13.75 & 13.76 & 4.986 & 4.992\tabularnewline
\hline 
 & $1/3$ & 10.39 & 10.40 & 13.93 & 13.93 & 4.922 & 4.925\tabularnewline
\hline 
 & $1/2$ & 10.84 & 10.83 & 14.32 & 14.33 & 4.816 & 4.818\tabularnewline
\hline 
 & $2/3$ & 11.40 & 11.41 & 14.79 & 14.81 & 4.719 & 4.725\tabularnewline
\hline 
 & $3/4$ & 11.68 & 11.70 & 15.07 & 15.09 & 4.672 & 4.678\tabularnewline
\hline 
 & $1$ & 12.92 & 12.95 & 16.27 & 16.31 & 4.518 & 4.527\tabularnewline
\hline 
 & AFE & 10.92 & 10.94 & 14.38 & 14.36 & 4.850 & 4.857\tabularnewline
\hline 
\end{tabular}
\end{table}

The convergence of transport coefficients for $1000$ Td with increasing
$l_{\mathrm{max}}$ is shown in Table \ref{tab:BechmarkConv}. Since
an even-order approximation is required for the appropriate boundary
conditions, the $l_{\mathrm{max}}$ are odd in our calculations. Clearly
the two-term approximation ($l_{\mathrm{max}}=1$) leads to an over-estimation
of the ionization rate, mean energy and flux drift velocity by approximately
$2\%$. Indeed six terms are required to achieve convergence to four
significant figures.

\begin{table}[H]
\protect\caption{Convergence of transport properties with $l_{\mathrm{max}}$ for PII
model (\ref{eq:testmodel}) at $1000$ Td and $Q=1/2$.\label{tab:BechmarkConv}}

\centering{}%
\begin{tabular}{|c|c|c|c|}
\hline 
 & $\alpha^{\mathrm{ion}}/n_{0}$ & $\epsilon$ & $W$\tabularnewline
$l_{\mathrm{max}}$ & ($10^{-15}$m$^{3}$s$^{-1}$) & (eV) & ($10^{5}$ms$^{-1}$)\tabularnewline
\hline 
\hline 
$1$ & 12.77  & 18.23 & 5.460\tabularnewline
\hline 
$3$ & 12.47  & 17.96 & 5.350\tabularnewline
\hline 
$5$ & 12.48 & 17.95  & 5.349\tabularnewline
\hline 
$7$ & 12.48  & 17.95 & 5.349\tabularnewline
\hline 
\end{tabular}
\end{table}

\begin{figure}[H]
\protect\caption{Variation of mean energy, $\epsilon$, with energy sharing fraction,
$Q$, for PII model (\ref{eq:testmodel}) at a reduced field of 800
Td.\label{fig:BenchmarkPositron800Td}}

\centering{}\includegraphics[scale=0.6]{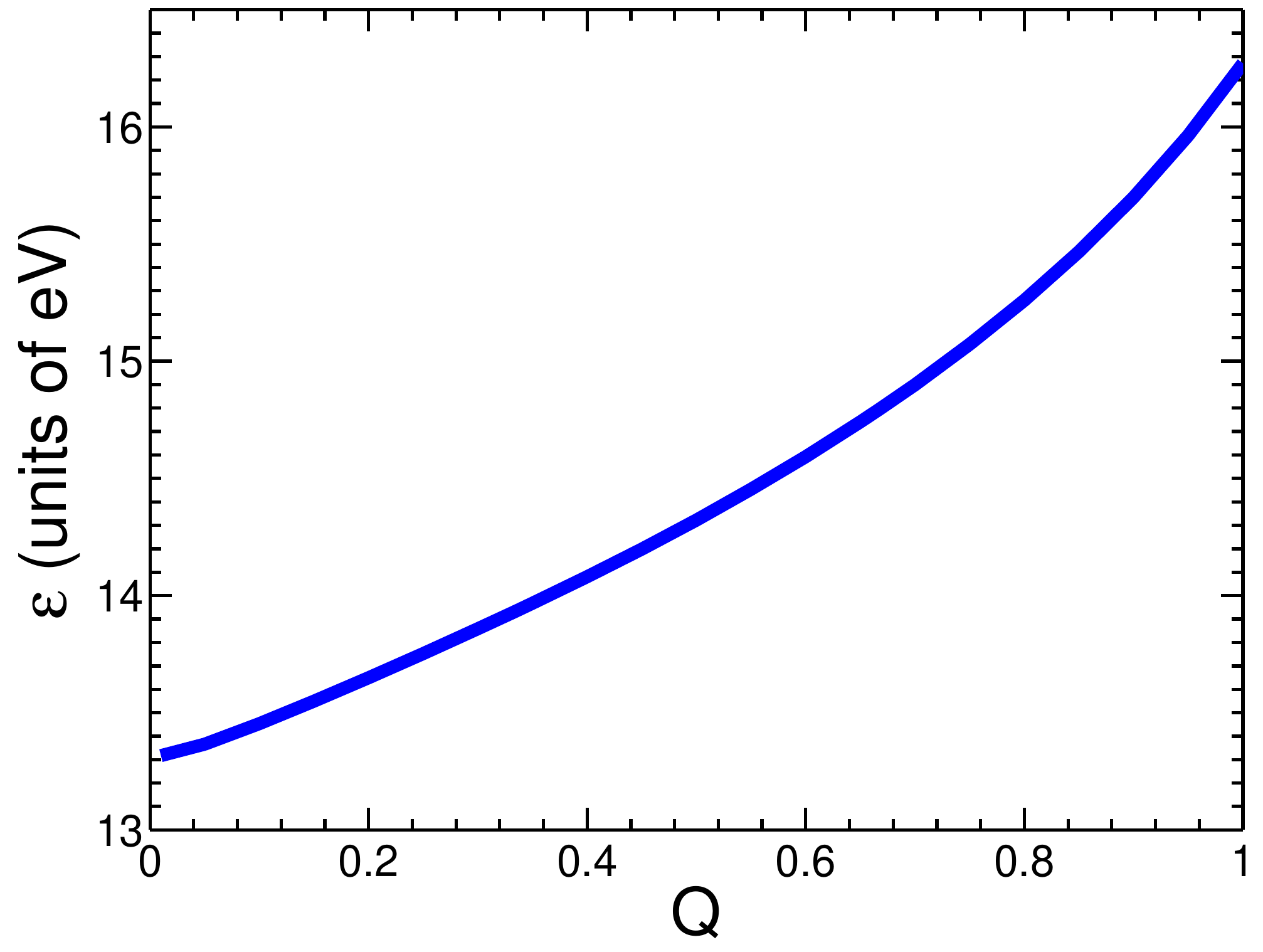}
\end{figure}

The variation of mean energy with $Q$ for PII at a reduced electric
field of $800$ Td is shown in Figure \ref{fig:BenchmarkPositron800Td}.
For PII, the mean energy of the positron swarm increases monotonically
with the energy-sharing fraction, $Q$. This behavior is to be expected,
as the ejected electron directly removes energy from the positron
swarm. The ionization collision frequency increases with energy in
model (\ref{eq:testmodel}), so that the greater the energy of the
swarm, the higher the rate of ionization collisions. Hence $\alpha^{\mathrm{ion}}/n_{0}$
also increases monotonically with $Q$. The flux drift velocity, $W$
in contrast, decreases with increasing $Q$. The effect of collisions
is to randomize the directions of the swarm particles, such that an
increase in the ionization rate decreases the average velocity of
the swarm. The transport properties for the `all fractions equiprobable'
(AFE) distribution are very similar to that of the equal-energy sharing
case.

\begin{figure}[H]
\protect\caption{Variation of mean energy, $\epsilon$, with energy sharing fraction,
$Q$, for EII model (\ref{eq:testmodel})at a reduced field of 800
Td.\label{fig:BenchmarkElectron800Td}}

\centering{}\includegraphics[scale=0.6]{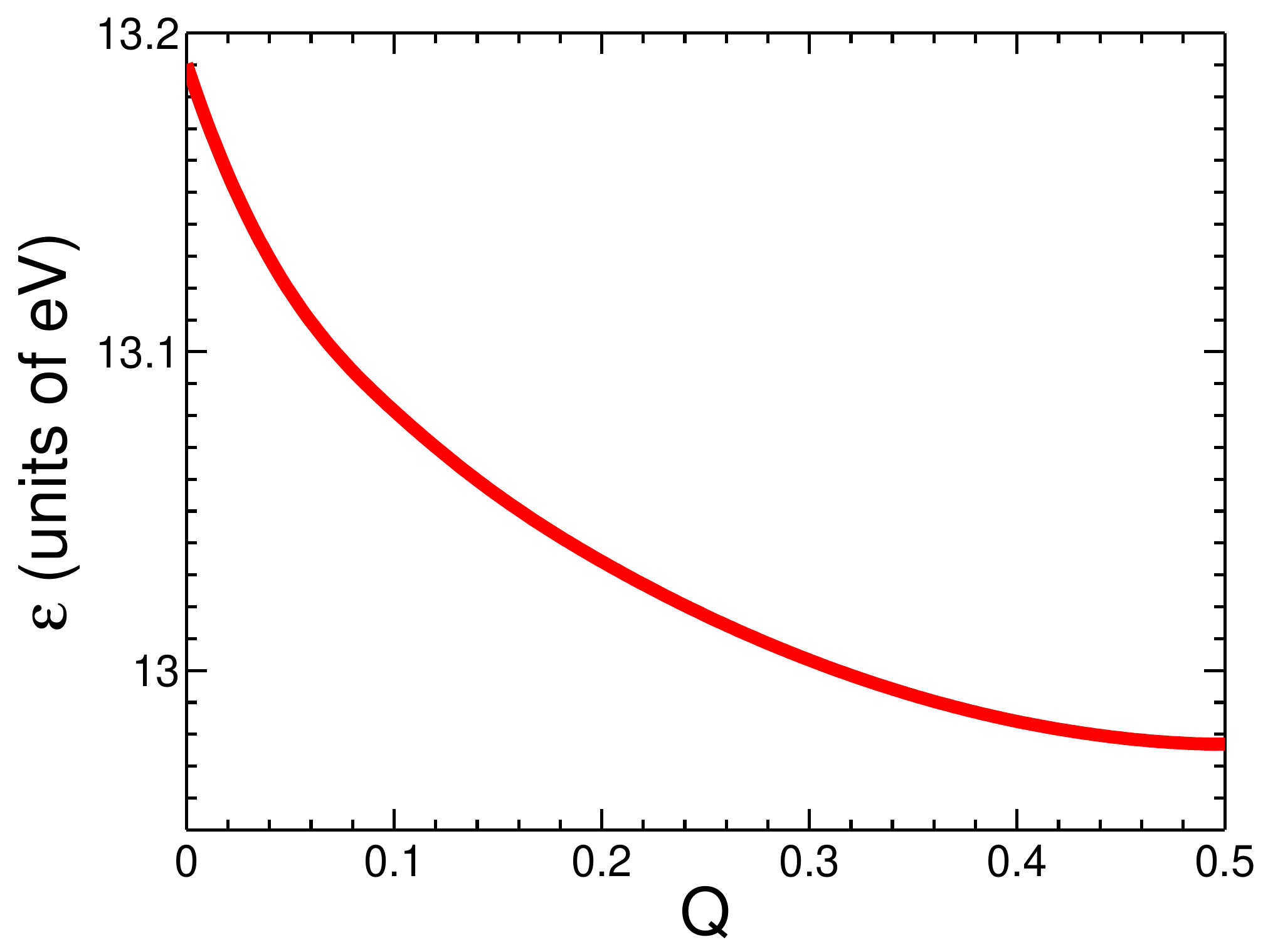}
\end{figure}

The variation of mean energy with $Q$ for EII at $800$ Td is shown
in Figure \ref{fig:BenchmarkElectron800Td}. The mean energy profile
is symmetrical about $Q=0.5$ due to the indistinguishability of post-collision
electrons, and for $800$ Td has a concave shape with a minimum value
corresponding to equal energy-sharing. It should be noted that, in
contrast to PII where the mean energy always increases with $Q$,
the exact nature of the EII mean energy profile depends on how the
distribution function samples the elastic, inelastic and ionization
cross sections. The variation in the transport properties for EII
with respect to $Q$ for the fields considered is small, suggesting
that EII is relatively insensitive to the exact nature of the energy-partitioning
for the model (\ref{eq:testmodel}). Ness and Makabe \cite{NessMaka00}
have shown that for EII in argon the choice of energy-sharing fraction
can in fact cause differences of $\sim25\%$, so that care must still
be taken when choosing the energy-partitioning function. 

The qualitative shape of the $Q$-dependence of the mean energy for
PII is insensitive to the reduced electric field, and the range of
values for a particular reduced field is considerably larger than
that for EII. In previous positron studies \cite{LoffWink96,Winketal02,Trunetal06,WhitRobs09},
PII has been treated as a standard excitation process. The current
results suggest that PII is particularly sensitive to the form of
the energy-partitioning and, if real-world PII differs significantly
from the model of pure scattering with excitation,  large errors can
result. To comment on this, we need to develop a realistic model of
PII energy-partitioning.

\subsection{Positron ionization energy-partitioning model}

We now wish to develop a model for post-ionization energy-partitioning
that captures the following basic physical behaviors:
\begin{enumerate}
\item For high impact energies, the positron ionization scattering cross
section approaches the electron ionization scattering cross section.
The first Born approximation \cite{BrauBrig86} is valid for high
impact energies and shows a heavy bias towards the case where the
scattered positron leaves the collision with almost all of the energy
which is available post-collision.
\item For impact energies near the ionization threshold, there is significant
correlation between the scattered positron and ejected electron. In
the Wannier theory \cite{Wann53} originally developed for near-threshold
EII, the repulsion between the two electrons cause them to emerge
with similar energies but in opposite directions. In terms of the
interaction potential between the two electrons, one may talk about
a Wannier \textquoteleft ridge\textquoteright{} upon which the system
is in an unstable equilibrium. Klar \cite{Klar81} was the first to
adapt Wannier's classical idea to PII. As in Wannier's theory the
energy is predicted to be shared equally, however now the positron
and electron emerge in similar directions due to the Coulomb attraction.
Ashley et al. \cite{Ashletal96} measured the positron ionization
cross-section in helium which they were able to accurately represent
by a power law, albeit different to that derived by Klar. Ihra et
al. \cite{Ihra97} extended the Wannier theory to be consistent with
both Klar and experiment. The success of these power law models justifies
the assumption of equal energy-sharing at near-threshold impact energies,
although recent experiments \cite{Arcietal05} suggest a slight asymmetry.
It should be noted that the positron and electron escape in similar
directions with similar energies and are highly correlated, so no
clear distinction between ionization and continuum state positronium
can be made \cite{CharHumb01}. 
\item Ionization at intermediate energies appears to be a combination of
the above two effects, i.e. a strong peak in the energy-sharing distribution
corresponding to the scattered positron leaving with all the available
energy, and a second peak occurring when the positron and electron
emerge with similar energy and direction and in a highly correlated
state. This feature has been shown in the studies of atomic hydrogen
by Brauner et al. \cite{Brauetal89} and measured experimentally in
H$_{2}$ by Laricchia and co-workers~\cite{KoveLari98,Arcietal05}.
\end{enumerate}
To capture simply the above three characteristics we propose a model
consisting of an exponentially decaying function, $g_{\mathrm{high}}(Q)$,
to represent the high impact energy ionization, and a rational polynomial
(sometimes called the Cauchy or Lorentz distribution), $g_{\mathrm{low}}(Q)$
centered around equal energy-sharing to represent the near-threshold
ionization i.e.,

\begin{align}
g_{\mathrm{high}}(Q) & =A_{\mathrm{high}}\exp(\beta_{\mathrm{high}}Q),\label{eq:ghigh}\\
g_{\mathrm{low}}(Q) & =A_{\mathrm{low}}\left[\beta_{\mathrm{low}}^{2}+\left(Q-0.5\right)^{2}\right]^{-1},\label{eq:glow}
\end{align}
where $Q$ is the fraction of the available energy, $A_{\mathrm{high}}$
and $A_{\mathrm{low}}$ are normalization constants, and $\beta_{\mathrm{high}}$
and $\beta_{\mathrm{low}}$ are free parameters to be fitted. An energy-fraction-partitioning
function which depends only on the impact energy and $Q$ can then
be constructed as  
\begin{align}
g(U',Q) & =w(U')g_{\mathrm{high}}(Q)+\left(1-w(U')\right)g_{\mathrm{low}}(Q),\label{eq:g}
\end{align}
where $w(U')$ is chosen as a hyperbolic tan function to transition
smoothly between $g_{\mathrm{high}}$ and $g_{\mathrm{low}}$, i.e.,
\begin{align}
w(U') & =\frac{1}{2}\left[1+\tanh\left(\gamma\frac{U'-U_{I}}{q}-\delta\right)\right],\label{eq:wtanh}
\end{align}
where $q$ is the elementary charge, and $\gamma$ and $\delta$ are
free parameters that control where and how sharp the transition is.
The relationship between energy-fraction-partitioning function, $g(U',Q)$,
and the energy-partitioning function, $P(U,U')$, used in equations
(\ref{eq:IonC}) and (\ref{eq:ionisation3}) is given simply by
\[
g(U',Q)Q=P(U,U')U.
\]

In the following subsections we shall investigate a test model with
reasonable values for the free parameters which can serve as a future
benchmark model, and then fit the energy-partitioning model to real
experimental H$_{2}$ data.

\subsubsection{Test model }

\begin{figure}[H]
\centering{}\protect\caption{Variation of the energy-fraction-partition function with impact energy,
relative to the ionization threshold, and energy sharing fraction,
$Q$, for parameters (\ref{eq:energysharemodel_parameters})\label{fig:BMenergyShare3D}}
 \includegraphics[scale=0.6]{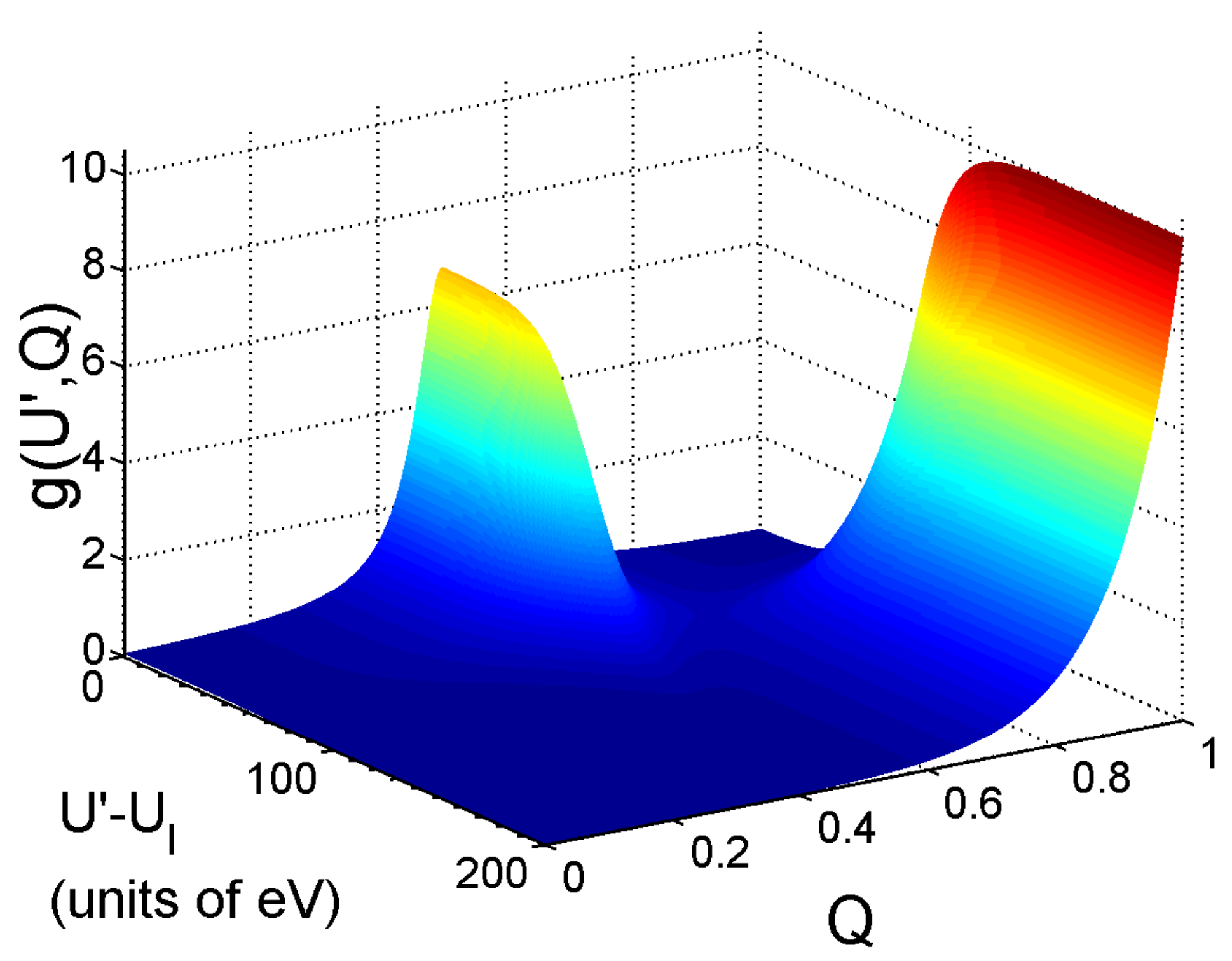}
\end{figure}

In this subsection we investigate the effect that the energy-partitioning
model (\ref{eq:ghigh})-(\ref{eq:wtanh}) has on positron transport
for a range of reduced electric field strengths. The parameters for
the energy-partitioning function are 
\begin{align}
\beta_{\mathrm{high}} & =10,\nonumber \\
\beta_{\mathrm{low}} & =0.05,\label{eq:energysharemodel_parameters}\\
\gamma & =0.05,\nonumber \\
\delta & =3.5,\nonumber 
\end{align}
with the same cross sections, neutral temperature and mass as the
model (\ref{eq:testmodel}). The energy-partition function for model
(\ref{eq:energysharemodel_parameters}) is displayed in Figure \ref{fig:BMenergyShare3D}.

Transport properties calculated via kinetic theory and Monte Carlo
are shown in Table \ref{tab:Benchmark3}. The kinetic theory and Monte
Carlo results agree to within $0.4\%$. Also included in the table
for $800$ Td and $5000$ Td are the swarm properties assuming the
energy-partitioning was replaced by only $g_{\mathrm{low}}$ or $g_{\mathrm{high}}$
respectively. At $800$ Td, the swarm properties for the full energy-partitioning
model are close to that which results from the inclusion of only $g_{\mathrm{low}}$,
which indicates that the distribution is generally sampling the even
energy-sharing part of the full energy-partitioning distribution.
At the higher field of $4000$ Td the swarm properties are now close
to those that come from allowing only $g_{\mathrm{high}}$ to have
an effect. As the field has increased, the distribution has shifted
from sampling mostly the even sharing region, to the region that is
heavily biased towards the positron getting large amounts of available
energy. 

\begin{table*}[tp]
\protect\caption{\label{tab:Benchmark3} Comparison of average ionization rate, $\alpha^{\mathrm{ion}}/n_{0}$,
mean energies, $\epsilon$, and flux drift velocities, $W$, for PII
for model (\ref{eq:energysharemodel_parameters}). The superscripts
$a$ and $b$ refer to $w(U)=0$ and $w(U)=1$ respectively. Columns
``Current'' correspond to the current kinetic theory calculations,
and columns ``MC'' are the results of Monte Carlo simulation.}

\centering{}%
\begin{tabular}{|c|c|>{\centering}p{1.2cm}|c|>{\centering}p{1.2cm}|c|>{\centering}p{1.2cm}|}
\hline 
$E/n_{0}$ & \multicolumn{2}{c|}{$\alpha^{\mathrm{ion}}/n_{0}$} & \multicolumn{2}{c|}{$\epsilon$} & \multicolumn{2}{c|}{$W$}\tabularnewline
(Td) & \multicolumn{2}{c|}{($10^{-15}$m$^{3}$s$^{-1}$)} & \multicolumn{2}{c|}{(eV)} & \multicolumn{2}{c|}{($10^{5}$ms$^{-1}$)}\tabularnewline
\hline 
 & Current & MC & Current & MC & Current & MC\tabularnewline
\hline 
\hline 
$800$ & 10.92 & 10.90 & 14.40 & 14.37 & 4.810 & 4.814\tabularnewline
\hline 
$800^{a}$ & 10.86 & 10.85 & 14.35 & 14.32 & 4.816 & 4.820\tabularnewline
\hline 
$800^{b}$ & 12.48 & 12.37 & 15.82 & 15.70 & 4.555 & 4.585\tabularnewline
\hline 
$1600$ & 26.29 & 26.26 & 34.12 & 34.04 & 6.331 & 6.348\tabularnewline
\hline 
$2400$ & 40.97 & 40.88 & 65.56 & 65.42 & 6.910 & 6.932\tabularnewline
\hline 
$3200$ & 53.97 & 53.85 & 104.1 & 103.9 & 7.201 & 7.229\tabularnewline
\hline 
$4000$ & 64.95 & 64.90 & 144.8 & 145.0 & 7.491 & 7.517\tabularnewline
\hline 
$4000^{a}$ & 49.18 & 49.11 & 86.49 & 86.52 & 9.509 & 9.527\tabularnewline
\hline 
$4000^{b}$ & 66.96 & 66.64 & 149.5 & 149.2 & 7.150 & 7.178\tabularnewline
\hline 
\end{tabular}
\end{table*}

\subsubsection{Model for positron-impact ionization in H$_{2}$}

\begin{figure}[H]
\protect\caption{\label{fig:H2energysharefit}Differential PII cross section for an
impact energy of 100 eV, as a function of the energy sharing fraction,
$Q$. KL 1998 is the experimental data of Kover and Laricchia \cite{KoveLari98}
for the triply differential cross section for an impact energy of
100 eV and ejection angle of $0^{\circ}$. The model fit has been
calculated with the parameters (\ref{eq:H2Partition}) and by assuming
that the triply differential cross section is the same at all ejection
angles. }

\centering{}\includegraphics[scale=0.6]{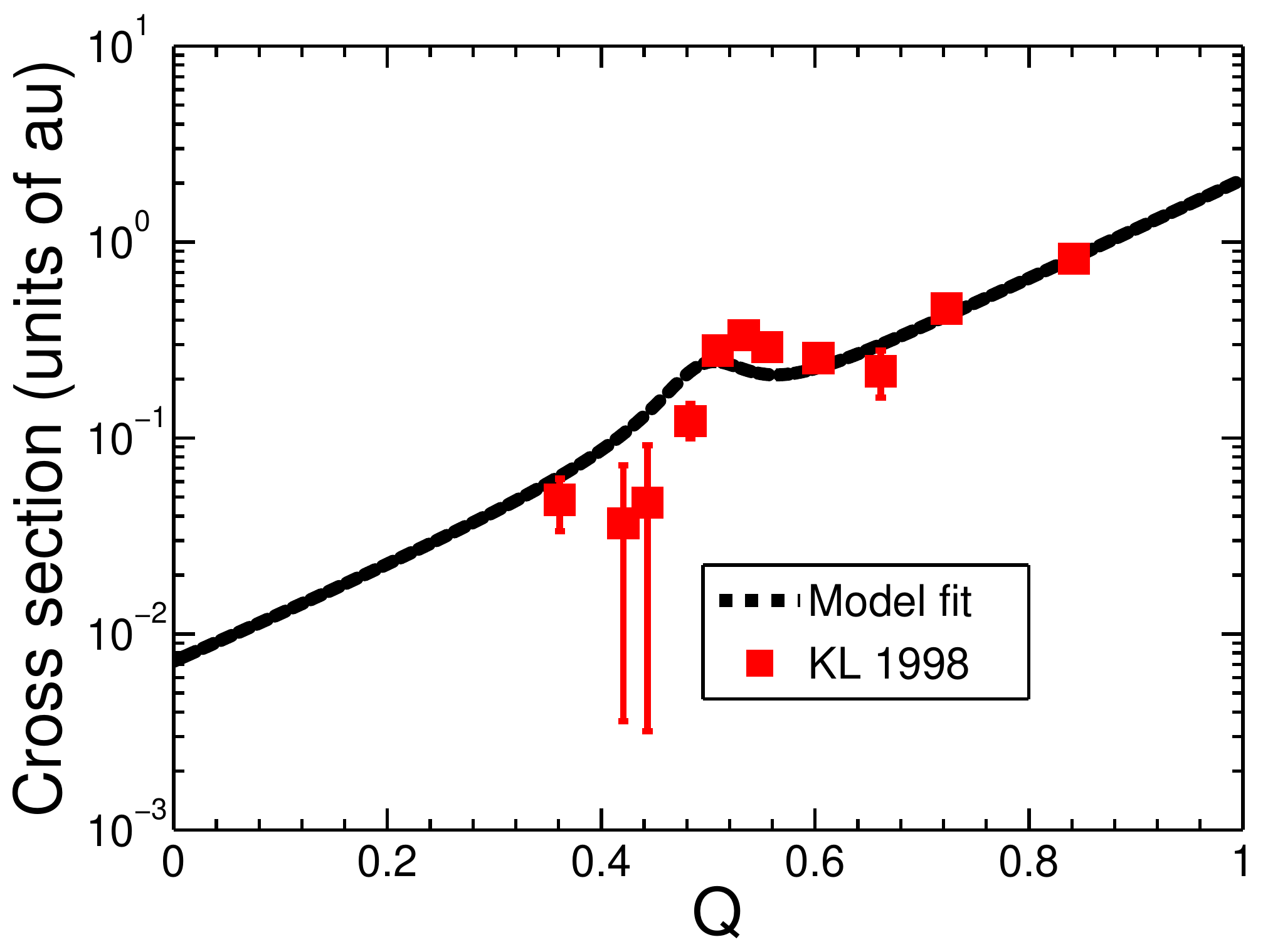}
\end{figure}

Laricchia and co-workers \cite{KoveLari98,Arcietal05} have measured
experimentally the energy-sharing of post-ionization species for PII
for a specific impact energy and angle. Their results for ionization
by a $100$ eV positron, where both the positron and electron emerge
at the same angle of $0$ degrees, are included in Figure \ref{fig:H2energysharefit}.
It is evident that there is a bias towards the positron getting all
or large amounts of the available energy, with a secondary peak close
to equal energy-sharing due to electron-positron correlation effects.
Our model predicts that this peak should occur at exactly $Q=0.5$,
but experiments show that there is a slight energy-sharing asymmetry
in positron ionization, such that the peak actually occurs at $Q>0.5$
\cite{Arcietal05}. A more sophisticated energy-partitioning model
will need to take this effect into account. We have performed a non-linear
least squares calculation to fit the free parameters of model (\ref{eq:ghigh})-(\ref{eq:wtanh})
to the experimental data, which were determined to be,

\begin{align}
\beta_{\mathrm{high}} & =5.88,\nonumber \\
\beta_{\mathrm{low}} & =0.0468,\label{eq:H2Partition}\\
\gamma & =0.0584,\nonumber \\
\delta & =3.45.\nonumber 
\end{align}
The fitted profile is shown in Figure \ref{fig:H2energysharefit}
and qualitatively reproduces the main features of the experiment.
It should be noted that at the $0$ degree scattering angle the secondary
peak is particularly dominant, and if one were to average the triple
differential cross section over all angles, a similar form with a
reduced secondary peak would result. Due to the lack of experimental
data at a variety of angles, we will assume that the angle-integrated
cross section has the exact same shape as the $0^{\circ}$ angle cross
section for the purpose of this paper, which will have the effect
of exaggerating the equal energy-sharing part of the full energy-sharing
distribution. The parameters in equation (\ref{eq:H2Partition}) have
been chosen to ensure a smooth transition between $g_{low}$ and $g_{high}$
while ensuring that the relative weights give the fit to experiment
for an impact energy of $100$ eV. The full, three dimensional energy-sharing
distribution is qualitatively similar to Figure \ref{fig:BMenergyShare3D}. 

\subsection{Positrons in molecular Hydrogen }

In the previous subsection, a model for the post-ionization energy-sharing
for PII from H$_{2}$ was proposed. In this subsection, the effect
of the energy sharing on transport properties is investigated for
PII in rarefied H$_{2}$. The set of H$_{2}$ cross sections employed
are those compiled in \cite{Banketal12b,Petretal13} and using the
elastic cross section of %
\footnote{M. Zammit in private communication\label{fn:Zammit}%
} calculated with a convergent-close-coupling formalism \cite{Zammetal13}
up to $1000$ eV, extrapolating where necessary. It is clear that
the ionization process, which turns on at $15.4$ eV is particularly
important, and dominates at energies above $50$ eV .

\begin{figure}[H]
\protect\caption{Cross section set for positron scattering in H$_{2}$. References
are given in text.\label{fig:H2-Cross-sections}}

\centering{}\includegraphics[scale=0.6]{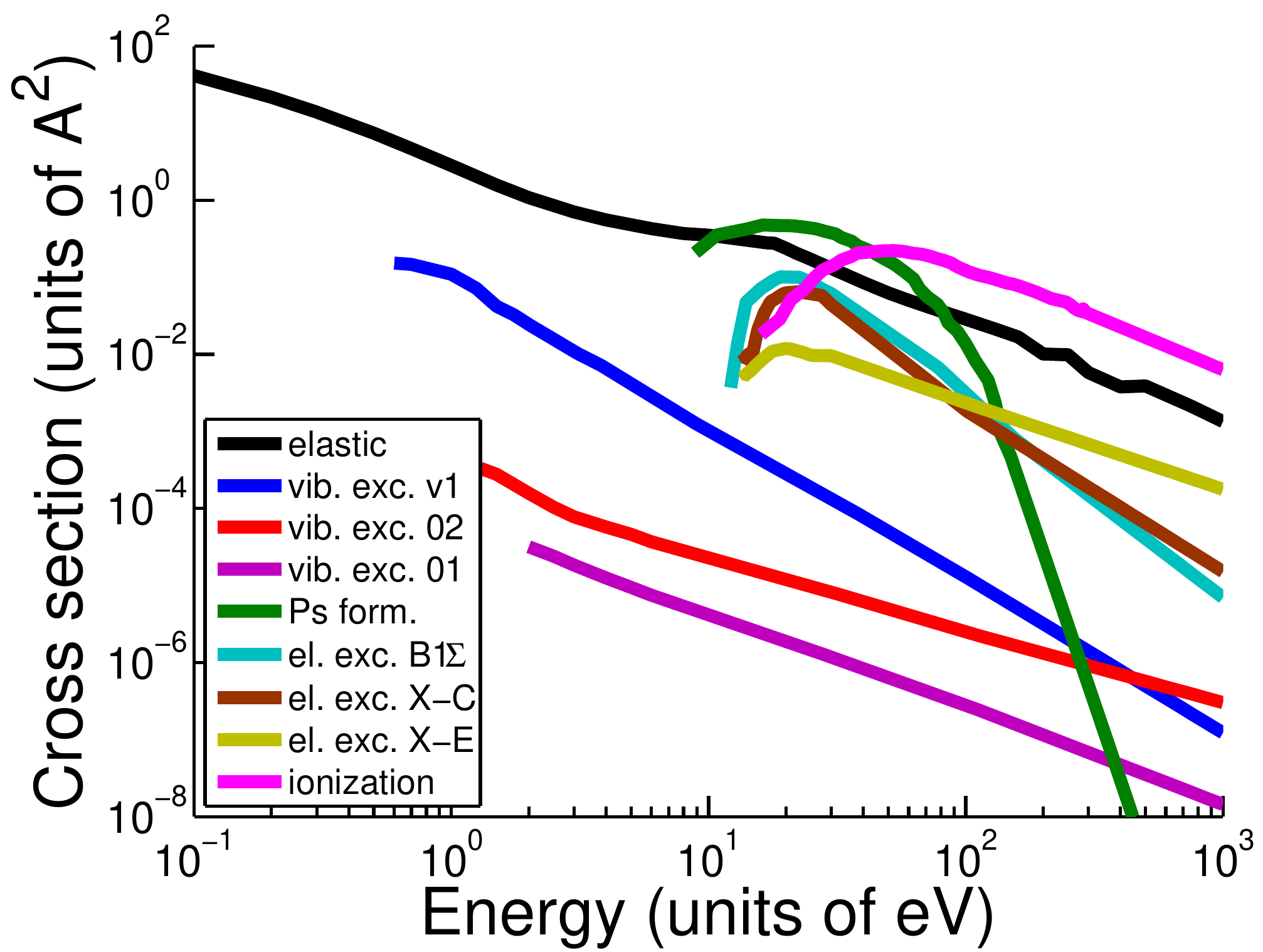}
\end{figure}

\begin{figure}[H]
\protect\caption{Mean energy temporal relaxation of a positron swarm in H$_{2}$ at
293 K. The initial source distribution is uniform in speed space up
to $1000$ eV. The H$_{2}$ model ionisation parameters are given
in equation~(\ref{eq:H2Partition}) and are compared with constant
energy sharing fractions of $Q=0.5$ and $Q=1.0$. \label{fig:H2-Therm}}

\centering{}\includegraphics[scale=0.6]{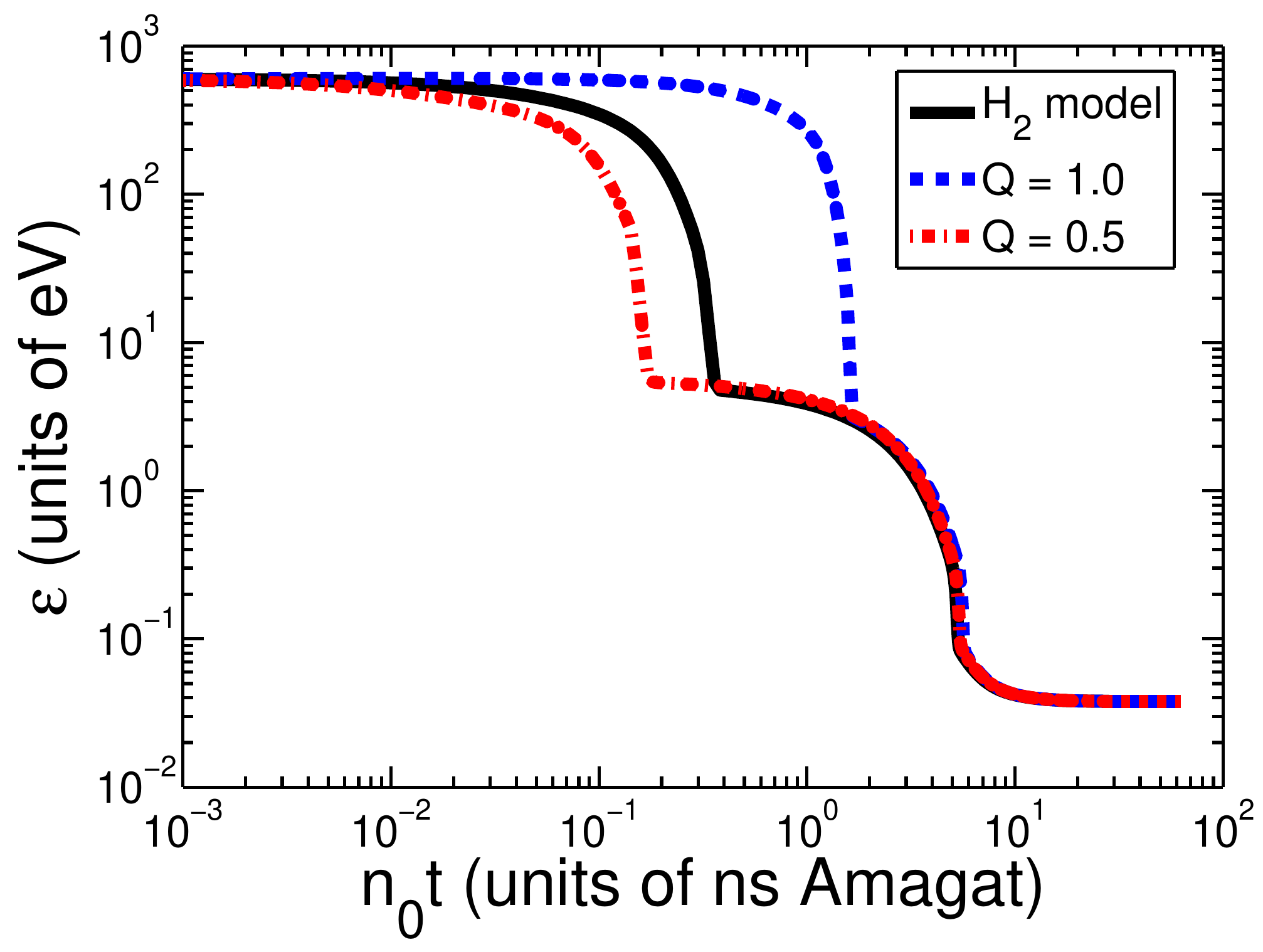}
\end{figure}

In order to assess the importance of energy-partitioning on ionization
we investigated the time-dependence of the mean energy for a source
of positrons in H$_{2}$ gas at $293$ K, as they relax to thermal
equilibrium in the absence of an electric field. The source distribution
is chosen to be uniform in velocity space up to the $1000$ eV cutoff,
which is equivalent to an initial mean energy of $600$ eV. The thermalisation
profiles for the energy-partitioning model (\ref{eq:H2Partition}),
and using the PII collision operator with $Q=0.5$ (equal energy-sharing),
and $Q=1$ (standard excitation form) are shown in Figure \ref{fig:H2-Therm}.
There are two distinct regions of rapid relaxation, one due to ionization
at high energies and one due to the vibrational modes at lower energies.
The first occurs on time scales of between $0.1$ and $2$ ns Amagat,
while the second at about $5$ ns Amagat, which shows that the relaxation
due to inelastic collisions is very rapid. While in the ionization-dominated
region, the three profiles show significant differences in mean energy
of up to an order of magnitude. The profile corresponding to $Q=1.0$
has the highest mean energy since the positron loses the least amount
of energy during an ionization collision in that limit. It takes significantly
longer to relax until the positron energies fall below the ionization
region, and thus they will experience more ionization collisions.
The $Q=0.5$ profile shows the lowest mean energy since the ejected
electron removes large amounts of energy from the swarm, and exits
the ionization region quickest. The ``real'' H$_{2}$ model profile
sits between the even energy-sharing and standard excitation profiles
as expected, since it is essentially a mixture of the two. At lower
energies, once ionization collisions become insignificant, all three
energy partitioning profiles coalesce, resulting in essentially the
same total thermalisation times. 

Although the total thermalisation time is essentially insensitive
to the form of the ionization energy-partitioning, the large differences
in mean energies in the ionization-dominated region can have other
important effects. In a space-dependent situation, the higher mean
energies can allow the positron to travel larger distances during
thermalisation. This is important to PET simulations since the resolution
of PET images is dependent on the distances traveled between positron
emission and annihilation \cite{CzerPhel02}. Similarly, the higher
the mean energy, the longer the positron swarm can significantly sample
the ionization cross section, and hence the more secondary electrons
that are created via PII. It is the secondary electrons created in
the human body during PET scans that can cause DNA damage \cite{Sanc03,Sanc05,Sanc08,Sanc09}.
Furthermore, the exact energy profile of the secondary electrons will
be dependent on the form of the PII energy-partitioning. 

\section{Conclusion\label{sec:Conclusion}}

Ionization by positron impact is a fundamentally different process
than ionization by electron impact. Applications such as PET demand
increasingly accurate models for positron transport, so it is important
to be able to describe the ionization process in detail. To this end,
a kinetic theory model with a general PII collision operator has been
developed for the first time. The key feature of the ionization collision
operator is the energy-partition function, which controls how the
available energy is shared between the post-collision constituents. 

The kinetic theory results were compared against a Monte Carlo simulation
for a simple test model (\ref{eq:testmodel}), which may serve as
a new benchmark for ionization. The transport properties calculated
differed between the two approaches by less than $0.6\%$ over a range
of reduced electric fields and available energy fractions, which is
within their respective uncertainties. The sensitivity of the transport
properties to the energy-sharing fraction $Q$ for PII was shown to
be significant, and much greater than that of EII. Thus large errors
can result in real-world applications if PII is not treated carefully.

A simple energy-partition function was developed to capture qualitatively
the underlying physics of PII. At high impact energies, the scattered
positron leaves the collision with almost all of the available energy,
while at near-threshold impact energies the Wannier theory \cite{Wann53}
suggests that the both the scattered positron and ejected electron
share approximately half of the available energy. In reality, there
is a slight energy-sharing asymmetry in near-threshold positron ionization
\cite{Arcietal05} and a more sophisticated energy-partitioning model
will need to take this asymmetry into account. The model parameters
were fit to the experimental results of Kover and Laricchia \cite{KoveLari98}
for positrons in H$_{2}$ with good qualitative agreement. 

Using the newly constructed H$_{2}$ energy-partitioning function,
we investigated the temporal relaxation of a positron swarm from a
high energy source ($600$ eV) to thermalisation at room temperature,
and compared the equal-energy sharing model with the common approach
of treating the PII as a standard excitation process. In the ionization-dominated
region there can be more than an order of magnitude in difference
in the mean energy profiles, and hence the choice of energy-partition
function has a significant effect on the number of ionization collisions
and the energy distribution of the secondary electrons created, which
is particularly important for radiation damage modeling \cite{Sanc08}.
Our modeling also suggests that the spatial relaxation will be sensitive
to the energy-partitioning, which is a topic to be further investigated.
\begin{acknowledgments}
This work was supported under the Australian Research Council's (ARC)
Centre of Excellence Discovery programs. The authors would like to
thank Prof. I. Bray and M. Zammit for helpful communications and cross
section calculations in H$_{2}$. S. Dujko acknowledges support from
MPNTRRS projects OI171037 and III41011.
\end{acknowledgments}
\bibliographystyle{apsrev}
\bibliography{PositronIonization}

\appendix

\section{Derivation of positron-impact ionization operator\label{sub:DerivationIonisation}}

The case of EII has been treated by Ness \cite{Ness85}, and we follow
this work closely to derive the PII collision operator. For simplicity,
we consider one ionization process with a neutral in the ground state,
but the generalization is straightforward. To derive the collision
operator we will consider the scattering of positrons into and out
of an element of phase space, $d\mathbf{r}d\mathbf{v}$.

Let us consider a beam of positrons incident upon the background neutrals
which are at rest. The flux of incident positrons, $\mathbf{I}$,
in $d\mathbf{r}d\mathbf{v}$ is
\begin{align*}
\mathbf{I} & =\mathbf{v}f\left(\mathbf{r},\mathbf{v},t\right)d\mathbf{v}.
\end{align*}
If $\sigma^{\mathrm{ion}}(v)$ is the the total ionization cross section
for an incoming positron of speed $v$, then the number of ionization
collisions in $d\mathbf{r}d\mathbf{v}$ per unit time per neutral
is,
\begin{align*}
I\sigma^{\mathrm{ion}}(v) & =vf(\mathbf{r},\mathbf{v},t)\sigma^{\mathrm{ion}}\left(v\right)d\mathbf{v},
\end{align*}
and hence, the total rate of positrons scattered out of the element
$d\mathbf{r}d\mathbf{v}$ for $n_{0}$ neutral particles due to ionization
is

\begin{align}
J_{\mathrm{out}}^{\mathrm{ion}}(f)d\mathbf{r}d\mathbf{v} & =n_{0}v\sigma^{\mathrm{ion}}(v)f(\mathbf{r},\mathbf{v},t)d\mathbf{r}d\mathbf{v}.\label{eq:Jion_out}
\end{align}

In EII, either the primary or ejected electrons (which are indistinguishable)
from an ionization event somewhere else in phase space may be scattered
into the element $d\mathbf{r}d\mathbf{v}.$ Since one can distinguish
between electrons and positrons, the PII equivalent is simpler. Let
us consider a new element of phase space with the same configuration
space location but new velocity space location, i.e., $d\mathbf{r}d\mathbf{v}'$.
Similar to (\ref{eq:Jion_out}), the total number of PII in $d\mathbf{r}d\mathbf{v}'$
per unit time is
\begin{align}
n_{0}v'\sigma^{\mathrm{ion}}(v')f(\mathbf{r},\mathbf{v}',t)d\mathbf{r}d\mathbf{v'}.\label{eq:ionizations1}
\end{align}
The momentum post-ionization is shared between the scattered positron
and the ejected electron. We define a quantity $B(\mathbf{v},\mathbf{v}'),$
such that $B(\mathbf{v},\mathbf{v}')d\mathbf{v}$ is the probability
of the positron having a velocity between $\mathbf{v}$ and $\mathbf{v}+d\mathbf{v}$
after ionization, given that the incident positron has velocity $\mathbf{v}'$.
Assuming the neutral particle remains a bystander at rest during the
process (to zeroth order in the mass ratio, $m/m_{0}$), then by conservation
of momentum,
\begin{align*}
\mathbf{v}' & =\mathbf{v}+\mathbf{\bar{v}},
\end{align*}
where $\mathbf{\bar{v}}$ is the velocity of the ejected electron.
It follows from equation (\ref{eq:ionizations1}) and the definition
of $B(\mathbf{v},\mathbf{v}')$ that the number of positrons that
enter $d\mathbf{r}d\mathbf{v}$ per unit time due to an ionization
event in $d\mathbf{r}d\mathbf{v'}$ is
\begin{align*}
n_{0}v'\sigma^{\mathrm{ion}}(v')f(\mathbf{r},\mathbf{v}',t)B(\mathbf{v},\mathbf{v}')d\mathbf{v}d\mathbf{r}d\mathbf{v'}.
\end{align*}
Integrating over all possible incident velocities thus yields the
total rate of positrons scattered into $d\mathbf{r}d\mathbf{v}$ due
to PII, i.e., 
\begin{align*}
J_{\mathrm{into}}^{\mathrm{ion}}(f)d\mathbf{r}d\mathbf{v} & =n_{0}d\mathbf{r}d\mathbf{v}\int v'\sigma^{\mathrm{ion}}(v')f(\mathbf{r},\mathbf{v}',t)B(\mathbf{v},\mathbf{v}')d\mathbf{v'}.
\end{align*}

The total PII collision operator is then the difference in the rates
of positrons scattered into and out of the element $d\mathbf{r}d\mathbf{v}$,
i.e., $J^{\mathrm{ion}}=J_{\mathrm{out}}^{\mathrm{ion}}-J_{\mathrm{into}}^{\mathrm{ion}},$

\begin{align}
J^{\mathrm{ion}}(f) & =n_{0}v\sigma^{\mathrm{ion}}(v)f(\mathbf{r},\mathbf{v},t)-n_{0}\int v'\sigma^{\mathrm{ion}}(v')B(\mathbf{v},\mathbf{v}')f(\mathbf{r},\mathbf{v}',t)d\mathbf{v'}.\label{eq:totalionization1}
\end{align}
If we assume central forces, then the scattering cross section and
partition function are dependent only on the magnitudes of the pre-
and post-collision velocities, and the angle between them, i.e., $v$,
$v'$ and $\hat{\mathbf{v}}\cdot\hat{\mathbf{v}}'$. We may then further
define a differential scattering cross section for ionization, $\sigma^{\mathrm{ion}}\left(v,v';\hat{\mathbf{v}}\cdot\hat{\mathbf{v}}'\right)$,
such that $\sigma^{\mathrm{ion}}\left(v,v';\hat{\mathbf{v}}\cdot\hat{\mathbf{v}}'\right)d\mathbf{v}$
is the number of positrons scattered into the range $d\mathbf{v}$
about \textbf{$\mathbf{v}$} due to incident electrons of velocity
$\mathbf{v}'$ divided by incident flux, 
\begin{align}
\sigma^{\mathrm{ion}}\left(v,v';\hat{\mathbf{v}}\cdot\hat{\mathbf{v}}'\right)d\mathbf{v} & =\sigma^{\mathrm{ion}}\left(v'\right)B\left(v,v';\hat{\mathbf{v}}\cdot\hat{\mathbf{v}}'\right)d\mathbf{v}.\label{eq:diffionXsect}
\end{align}
The partition function satisfies a normalization condition so that
\begin{align*}
\sigma^{\mathrm{ion}}(v') & =\int\sigma^{\mathrm{ion}}\left(v,v';\hat{\mathbf{v}}\cdot\hat{\mathbf{v}}'\right)d\mathbf{v}.
\end{align*}
Substituting equation (\ref{eq:diffionXsect}) into equation (\ref{eq:totalionization1})
gives the PII collision operator 
\begin{align*}
J^{\mathrm{ion}}(f) & =n_{0}v\sigma^{\mathrm{ion}}(v)f(\mathbf{v})-n_{0}\int v'\sigma^{\mathrm{ion}}\left(v,v';\hat{\mathbf{v}}\cdot\hat{\mathbf{v}}'\right)f(\mathbf{v}')d\mathbf{v'}.
\end{align*}
This operator is particle-number-conserving, i.e.
\begin{align*}
\int J^{\mathrm{ion}}(f)d\mathbf{v} & =\int n_{0}v\sigma^{\mathrm{ion}}(v)f(\mathbf{v})d\mathbf{v}-n_{0}\iint v'\sigma^{\mathrm{ion}}\left(v,v';\hat{\mathbf{v}}\cdot\hat{\mathbf{v}}'\right)f(\mathbf{v}')d\mathbf{v'}d\mathbf{v}\\
 & =n_{0}\int v\sigma^{\mathrm{ion}}(v)f(\mathbf{v})d\mathbf{v}-n_{0}\int v'f(\mathbf{v}')d\mathbf{v'}\int\sigma^{\mathrm{ion}}\left(v,v';\hat{\mathbf{v}}\cdot\hat{\mathbf{v}}'\right)d\mathbf{v}\\
 & =n_{0}\int v\sigma^{\mathrm{ion}}(v)f(\mathbf{v})d\mathbf{v}-n_{0}\int v'\sigma^{\mathrm{ion}}(v')f(\mathbf{v}')d\mathbf{v'}\\
 & =0,
\end{align*}
as required.

\subsubsection{Legendre decomposition}

For central scattering forces the partition function can be decomposed
in terms of Legendre polynomials, i.e.,
\begin{align*}
B_{l}(v,v') & =2\pi\int_{-1}^{1}B(\mathbf{v},\mathbf{v}')P_{l}\left(\mu\right)d\mu,
\end{align*}
where $\mu=\hat{\mathbf{v}}\cdot\hat{\mathbf{v}}'$. For isotropic
scattering, $B_{l}(v,v')=0$ for $l\geq1$. Multiplying equation (\ref{eq:totalionization1})
by $P_{l}\left(\cos\chi\right)$, and integrating over all angles
leads to 
\begin{align}
J_{l}^{\mathrm{ion}}(f_{l}) & =n_{0}v\sigma^{\mathrm{ion}}(v)f_{l}(v)-\begin{cases}
n_{0}\int_{0}^{\infty}v'\sigma^{\mathrm{ion}}\left(v'\right)B_{0}(v,v')f_{0}(v')v'^{2}dv' & l=0,\\
0 & l\geq1.
\end{cases}\label{eq:ionisation2}
\end{align}
We now seek to represent equation (\ref{eq:ionisation2}) in terms
of energy rather than speed, i.e., $U=\frac{1}{2}mv^{2}$. The probability
of a positron having a speed in the range $v+dv$ after ionization,
for an incident positron of speed $v'$ is

\begin{align}
v^{2}dv\int B\left(v,v';\mathbf{\hat{v}}\cdot\mathbf{\hat{v}}'\right)d\mbox{\ensuremath{\mathbf{\hat{v}}}} & =B\left(v,v'\right)v^{2}dv,\nonumber \\
 & \equiv P\left(U,U'\right)dU,\label{eq:angleB2}
\end{align}
where $U$ and $U'$ are the post- and pre-collision positron energies
respectively, and now the right-hand-side term of equation (\ref{eq:angleB2})
represents the probability of a positron having an energy in the range
$U+dU$ after ionization for an incident positron of $U'$. The energy
partitioning function, $P(U,U')$, has the following properties:

\begin{align*}
P(U,U') & =0\ \ U'<U+U_{I}\\
\int_{0}^{U'-U_{I}}P(U,U')dU & =1\ \ U'\geq U+U_{I}.
\end{align*}

Finally, we can represent equation (\ref{eq:ionisation2}) in terms
of energy and the energy-partition function, $P(U,U')$, 

\begin{align}
J_{l}^{\mathrm{ion}}(f_{l}) & =n_{0}\sqrt{\frac{2U}{m}}\sigma^{\mathrm{ion}}(U)f_{l}(U)-\begin{cases}
\begin{array}{c}
n_{0}\sqrt{\frac{2}{mU}}\int_{0}^{\infty}U'\sigma^{\mathrm{ion}}\left(U'\right)P(U,U')f_{0}(U')dU'\\
0
\end{array} & \begin{array}{c}
l=0,\\
l\geq1,
\end{array}\end{cases}\label{eq:ionisation3}
\end{align}

\subsubsection{Modified Frost-Phelps operator}

If the scattered positron leaves the collision with an exact fraction,
$Q$, of the available energy, $U'-U_{I}$, where $U_{I}$ is the
threshold energy, then the energy-partition function has the form,
\begin{align*}
P(U,U') & =\delta\left(U-Q(U'-U_{I})\right),\\
 & =\frac{1}{Q}\delta\left(U'-\left(\frac{U}{Q}+U_{I}\right)\right),
\end{align*}
and the integral in equation (\ref{eq:ionisation3}) reduces to,
\begin{align}
J_{l}^{\mathrm{ion}}(f_{l}) & =\nu^{\mathrm{ion}}(U)f_{l}(U)-\begin{cases}
\begin{array}{c}
\frac{1}{Q}\frac{\left(\frac{U}{Q}+U_{I}\right){}^{\frac{1}{2}}}{U^{\frac{1}{2}}}\nu^{\mathrm{ion}}\left(\frac{U}{Q}+U_{I}\right)f_{0}\left(\frac{U}{Q}+U_{I}\right)\\
0
\end{array} & \begin{array}{c}
l=0,\\
l\geq1,
\end{array}\end{cases}\label{eq:modfrostphelps}
\end{align}
where $\nu^{\mathrm{ion}}(U)=n_{0}\sqrt{\frac{2U}{m}}\sigma^{\mathrm{ion}}(U)$
is the ionization collision frequency. Equation (\ref{eq:modfrostphelps})
can be considered a `modified Frost-Phelps' operator. A similar result
for EII was given in \cite{SigeWink97}. In the case where the positron
gets all of the available energy, i.e., $Q=1$, equation (\ref{eq:modfrostphelps})
reduces to the standard Frost-Phelps operator, (\ref{eq:inelastic}),
as required. Clearly, equation (\ref{eq:modfrostphelps}) breaks down
when $Q=0$.

\section{Electron impact ionization benchmarks\label{sub:Electron-Impact-Ionization}}

Transport coefficients for EII are given in Table \ref{tab:ElectronBenchmark},
in which they are compared to the results of Ness and Robson \cite{NessRobs86}.
Due to the indistinguishability of post-collision particles, the results
for $Q$ and $1-Q$ with respect to EII are identical, and so we consider
only $Q>0.5$. The modified Frost-Phelps form of the collision operator
(\ref{eq:modfrostphelps}) breaks down when $Q=0$, hence there is
no value given in Table \ref{tab:ElectronBenchmark} corresponding
to $Q=0$ and $Q=1$ (if one of the electrons gets the fraction $Q=1$
of the available energy, then the other receives $Q=0$ and the same
problem is encountered). The EII calculations using our kinetic theory
model agree closely with both our Monte Carlo simulations and the
kinetic theory approach in \cite{NessRobs86}. There are generally
differences of less than $0.6\%$ and $0.3\%$ in the ionization rate
and mean energy respectively, between the present kinetic theory results
and both the Monte Carlo simulation and \cite{Ness85} over the whole
range of reduced fields and energy sharing fractions, except for the
AFE case. An error is present in the AFE calculations of \cite{Ness85}.
Values for the AFE case have been re-calculated using a similar Burnett
function \cite{Kumaetal80} expansion to that of Ness and Robson (which
are included in Table \ref{tab:ElectronBenchmark} enclosed within
square brackets) which agree closely with our calculations. In reference
\cite{Ness85}, the bulk drift velocities are given, which must not
be confused with the flux drift velocity \cite{Sakaetal77,Tagaetal77,RobsNess86}.
The two types of transport coefficients can be significantly different
when there a non-conservative effects. Following \cite{RobsNess86}
we have solved the first level of spatially-inhomogeneous equations,
which come from a density gradient expansion \cite{Kumaetal80}, in
addition to equation (\ref{eq:BoltzLeg}) to determine the bulk drift
velocity. Both the flux and bulk drift velocities generally agree
to within $0.3\%$ between the three calculation methods over the
range of fields and energy-sharing fractions considered.

\begin{table}[H]
\protect\caption{Comparison of average ionization rate, $\alpha^{\mathrm{ion}}/n_{0}$,
mean energies, $\epsilon$, flux drift velocities, $W_{\mathrm{flux}}$,
and bulk drift velocities $W_{\mathrm{bulk}}$ for EII for model (\ref{eq:testmodel})
for different reduced fields $E/n_{0}$ and energy sharing fractions
$Q$. The first column lists the current kinetic theory calculations,
the second column lists the results of our Monte Carlo simulations,
and the third includes the kinetic theory calculations of Ness and
Robson \cite{NessRobs86}. The values enclosed in square brackets
have been performed using a similar Burnett function expansion to
that of Ness and Robson. $Q=\mathrm{AFE}$ corresponds to `all fractions
equiprobably'. Note: there was an error in the AFE case in the original
Ness and Robson work \cite{NessRobs86}.\label{tab:ElectronBenchmark}}

\centering{}%
\begin{tabular}{|c|c|c|>{\centering}p{1.2cm}|>{\centering}p{1.2cm}|c|c|>{\centering}p{1.2cm}|c|c|c|c|c|}
\hline 
$E/n_{0}$ &  & \multicolumn{3}{c|}{$\alpha^{\mathrm{ion}}/n_{0}$} & \multicolumn{3}{c|}{$\epsilon$} & \multicolumn{2}{c|}{$W_{\mathrm{flux}}$} & \multicolumn{3}{c|}{$W_{\mathrm{bulk}}$}\tabularnewline
(Td) & $Q$ & \multicolumn{3}{c|}{($10^{-15}$m$^{3}$s$^{-1}$)} & \multicolumn{3}{c|}{(eV)} & \multicolumn{2}{c|}{($10^{5}$ms$^{-1}$)} & \multicolumn{3}{c|}{($10^{5}$ms$^{-1}$)}\tabularnewline
\hline 
 &  & Current & MC & \cite{NessRobs86} & Current & MC & \cite{NessRobs86} & Current & \cite{NessRobs86} & Current & MC & \cite{NessRobs86}\tabularnewline
\hline 
\hline 
300 & 0 &  & 1.620 & 1.61 &  & 6.739 & 6.73 &  & 2.780 &  & 3.236 & 3.23\tabularnewline
\hline 
 & $1/4$ & 1.598 & 1.611 & 1.60 & 6.737 & 6.741 & 6.73 & 2.752 & 2.754 & 3.200 & 3.204 & 3.20\tabularnewline
\hline 
 & $1/3$ & 1.595 & 1.596 & 1.60 & 6.739 & 6.741 & 6.73 & 2.748 & 2.749 & 3.194 & 3.196 & 3.20\tabularnewline
\hline 
 & $1/2$ & 1.591 & 1.589 & 1.59 & 6.742 & 6.744 & 6.74 & 2.744 & 2.745 & 3.189 & 3.192 & 3.19\tabularnewline
\hline 
 & \multirow{2}{*}{AFE} & 1.600 & 1.606 & 1.51 & 6.733 & 6.746 & 6.75  & 2.756 & 2.755 & 3.198 & 3.206 & 3.19\tabularnewline
 &  &  &  & {[}1.60{]} &  &  & {[}6.73{]} &  &  &  &  & {[}3.21{]}\tabularnewline
\hline 
$500$ & $0$ &  & 4.643 & 4.68 &  & 9.009 & 8.99 &  & 3.920 &  & 4.752 & 4.74\tabularnewline
\hline 
 & $1/4$ & 4.504 & 4.515 & 4.51 & 9.007 & 9.007 & 9.01 & 3.835 & 3.839 & 4.632 & 4.644 & 4.63\tabularnewline
\hline 
 & $1/3$ & 4.482 & 4.492 & 4.49 & 9.013 & 9.023 & 9.01 & 3.823 & 3.822 & 4.617 & 4.617 & 4.62\tabularnewline
\hline 
 & $1/2$ & 4.464 & 4.452 & 4.47 & 9.017 & 9.028 & 9.02 & 3.814 & 3.816 & 4.604 & 4.606 & 4.61\tabularnewline
\hline 
 & \multirow{2}{*}{AFE} & 4.511 & 4.525 & 4.37 & 9.000 & 9.007 & 9.04 & 3.846 & 3.843 & 4.635 & 4.647 & 4.62\tabularnewline
 &  &  &  & {[}4.52{]} &  &  & {[}9.00{]} &  &  &  &  & {[}4.64{]}\tabularnewline
\hline 
$800${*} & $0$ &  & 9.736 & 9.62 &  & 13.17 & 13.21 &  & 5.112 &  & 6.284 & 6.25\tabularnewline
\hline 
 & $1/4$ & 9.413 & 9.422 & 9.41 & 13.01 & 13.02 & 13.01 & 4.953 & 4.957 & 6.108 & 6.118 & 6.11\tabularnewline
\hline 
 & $1/3$ & 9.357 & 9.372 & 9.37 & 12.99 & 12.99 & 12.99 & 4.933 & 4.936 & 6.090 & 6.092 & 6.09\tabularnewline
\hline 
 & $1/2$ & 9.320 & 9.339 & 9.33 & 12.97 & 12.98 & 12.97 & 4.919 & 4.922 & 6.079 & 6.095 & 6.08\tabularnewline
\hline 
 & \multirow{2}{*}{AFE} & 9.461 & 9.445 & 9.20  & 13.03 & 13.02 & 13.09  & 4.968 & 4.976 & 6.137 & 6.137 & 6.12\tabularnewline
 &  &  &  & {[}9.46{]} &  &  & {[}13.02{]} &  &  &  &  & {[}6.13{]}\tabularnewline
\hline 
\end{tabular}
\end{table}

\end{document}